\documentclass[%
 reprint,
superscriptaddress,
 amsmath,amssymb,
 aps,
prb,
]{revtex4-2}

\usepackage{graphicx}
\usepackage{subfigure}
\usepackage{float}
\usepackage{dcolumn}
\usepackage{bm}
\usepackage{rotating}
\usepackage{color}
\usepackage{ulem}

\usepackage[colorlinks, citecolor=blue,urlcolor=blue,linkcolor=blue]{hyperref}
\begin{document}

\preprint{APS/123-QED}

\title{Geometric origin of self-intersection points in non-Hermitian energy spectra} 

\author{Jinghui Pi}

\thanks{These authors contributed equally.\\\href{mailto:pijh14@gmail.com}{pijh14@gmail.com}}

 \affiliation{The Chinese University of Hong Kong Shenzhen Research Institute, 518057 Shenzhen, China}
 
\author{Chenyang Wang}%
\thanks{These authors contributed equally.\\\href{mailto:pijh14@gmail.com}{pijh14@gmail.com}}
\affiliation{State Key Laboratory of Low-Dimensional Quantum Physics, Department of Physics, Tsinghua University, Beijing 100084, China
}%


\author{Yong-Chun Liu}
\affiliation{State Key Laboratory of Low-Dimensional Quantum Physics, Department of Physics, Tsinghua University, Beijing 100084, China
}%
\affiliation{
 Frontier Science Center for Quantum Information, Beijing 100084, China
}%

\author{Yangqian Yan}
\email{yqyan@cuhk.edu.hk}
\affiliation{Department of Physics, The Chinese University of Hong Kong, Shatin, New Territories, Hong Kong, China
} 
\affiliation{The Chinese University of Hong Kong Shenzhen Research Institute, 518057 Shenzhen, China}

 %



\begin{abstract}
  Unlike Hermitian systems, non-Hermitian energy spectra under periodic boundary conditions can form closed loops in the complex energy plane, a phenomenon known as the point-gap topology. In this paper, we investigate the self-intersection points of such non-Hermitian energy spectra and reveal their geometric origins. We rigorously demonstrate that these self-intersection points result from the intersection of the auxiliary generalized Brillouin zone and the Brillouin zone in one-band systems, as confirmed by an extended Hatano-Nelson model. This finding is further generalized to multi-band systems, illustrated through a non-Hermitian Su-Schrieffer-Heeger model. Moreover, we address multiple self-intersection points and derive the geometric conditions for general $n$-fold self-intersection points. Our results enhance the fundamental understanding of generic non-Hermitian quantum systems and provide theoretical support for further experimental investigations of energy self-intersection points. 
\end{abstract}

\maketitle


\section{Introduction}
Non-Hermiticity emerges when a closed system couples with its environment \cite{El2018Non}. It has been realized in many experimental platforms, including optical systems with gain and loss \cite{Feng_2017, Longhi_2017, Ozawa_2019}, open systems with dissipation \cite{Rotter_2009}, and electron systems with finite-lifetime quasi-particles \cite{Yoshida_2018, Shen_2018, Yamamoto_2019}. A variety of phenomena beyond traditional paradigms have been revealed~\cite{Bender_2007,RevModPhys.87.61,RevModPhys.88.035002,ashida2020non,RevModPhys.93.015005,ding2022non,doi:10.1146/annurev-conmatphys-040521-033133,Yang_2024}, including the non-Hermitian skin effect (NHSE) \cite{PhysRevLett.116.133903,PhysRevLett.121.086803,PhysRevLett.121.136802,PhysRevLett.121.026808,NHSE_2018,PhysRevB.99.201103}, which refers to the boundary localization of a large portion of eigenstates. The NHSE renders a high sensitivity of spectra to the boundary conditions, thereby breaking conventional bulk-edge correspondence \cite{Xiao2020Non,helbig2020generalized,doi:10.1073/pnas.2010580117}.
A central concept related to the NHSE is the generalized Brillouin zone (GBZ) \cite{PhysRevLett.121.086803,PhysRevLett.121.136802}. By replacing the Brillouin zone (BZ) with the GBZ, one can establish the non-Bloch band theory and reshape the bulk-edge correspondence \cite{PhysRevLett.123.066404, PhysRevResearch.1.023013,PhysRevLett.123.016805, PhysRevLett.125.126402,PhysRevLett.125.186802,PhysRevB.101.195147, PhysRevLett.126.216405, Zhang_2022universal, PhysRevLett.128.223903,PhysRevX.14.021011,HU2024,PhysRevB.110.075411,PhysRevLett.132.086502}. An efficient analytical and numerical approach to obtaining the GBZ is the auxiliary GBZ (aGBZ) method \cite{PhysRevLett.122.250201,PhysRevLett.125.226402}, which also manifests the critical NHSE \cite{Li_2020,PhysRevResearch.2.043167, PhysRevB.104.165117, PhysRevB.107.155430}.

The topological origin of the NHSE is the non-trivial non-Hermitian point-gap \cite{PhysRevLett.125.126402, PhysRevX.9.041015,PhysRevLett.124.086801,PhysRevLett.132.136401,PhysRevLett.124.056802}. Specifically, the energy spectra under periodic boundary conditions (PBC) form a closed loop enclosing a finite area when systems feature the NHSE under the open boundary conditions (OBC). Such PBC energy spectra can be categorized into two types: simple closed curves with a single connected interior region, and self-intersecting closed curves that partition the interior into distinct sub-regions. When systems exhibit the bipolar skin effect \cite{PhysRevLett.123.246801,PhysRevB.109.214311}, the PBC energy spectra must have self-intersection points and belong to the latter type of curves. However, if all skin modes are localized at one boundary, the PBC energy spectra may also display self-intersection points \cite{PhysRevB.106.235411}. Therefore, it remains unclear how these self-intersection points arise in the PBC energy spectra.

In this work, we examine self-intersection points within general one-dimensional non-Hermitian energy spectra and offer a comprehensive explanation for the geometric origins of these points. Notably, as non-Hermitian degeneracies, energy self-intersection points are distinct from exceptional points \cite{Berry2004,PhysRevE.69.056216,Rotter_2009,PhysRevLett.118.040401,Xiong_2018,PhysRevB.110.L201104, PhysRevLett.128.010402,qin2024nonlinear}, where eigenstates are coalesced. For convenience, we review the one-band theory in one dimensional non-Hermitian systems and introduce the concept of the energy winding number, the GBZ, and the aGBZ. We uncover that, for one-band systems, the PBC energy spectra exhibit self-intersection points if and only if the corresponding aGBZ curve intersects with the unit circle (the BZ curve), and vice versa. This finding is rigorously demonstrated through a topological method and confirmed by numerical investigations of an extended Hatano-Nelson model. Furthermore, we extend our analysis to the multi-band cases and illustrate the results through a non-Hermitian Su-Schrieffer-Heeger (SSH) model. Additionally, we discuss the multiple self-intersection points with multiplicities greater than two, deriving the geometric conditions for general $n$-fold self-intersection points. Our findings advance the foundational understanding of the spectra of generic non-Hermitian quantum systems, and offer theoretical justification for further experimental explorations of energy self-intersection points.
    
The rest of the paper is organized as follows. We provide a brief review of the fundamental concept of the one-band non-Hermitian systems in Sec.~\ref{sec2}, including the energy winding number, the GBZ, and the aGBZ. In Sec.~\ref{sec3}, we rigorously prove that self-intersecting points in one-band PBC energy spectra originate from the intersection of the aGBZ and the BZ. The above one-band conclusion is extended to the multi-band cases in Sec.~\ref{sec4}, and the geometric conditions for general $n$-fold self-intersection points is established in Sec.~\ref{sec5}. Finally, a summary and discussion are presented in Sec.~\ref{sec6}.

\section{Review of one-band non-Hermitian systems}\label{sec2}
We begin with a general one-band tight-binding model in one dimension. The real space Hamiltonian reads
\begin{equation}
 H=\sum_{i,j}t_{j-i}\left\vert i\right\rangle \left\langle j\right\vert , 
 \label{eq1}
\end{equation}
where $i,j$ are the position indices and the hopping amplitudes, $t_{j-i}$,
depend solely on the spatial distance $j-i$ and have finite range with left
(right) hopping range $M$ ($N$). Applying the Fourier transformation of
hopping amplitudes, we obtain the Bloch Hamiltonian,
\begin{equation}
h\left(  k\right)  =\sum_{n=-M}^{N}t_{n}\left(  e^{ik}\right)  ^{n}.
\label{eq2}
\end{equation}
For the periodic boundary condition, the Bloch
phase factor $e^{ik}$ moves along the unit circle on the complex plane with $0\leqslant k<2\pi$,. In the
Hermitian case, where $t_{n}=t_{-n}^{\ast}$, the  spectrum of $h\left(
k\right)  $ lies on the real axis and thus encloses zero area. However, for the
non-Hermitian systems with $t_{n}\neq t_{-n}^{\ast}$, the  spectrum of
$h\left(  k\right)  $ can form a closed curve in the complex energy plane and
encloses a finite area. Such a distinct topological behavior is
characterized by the winding number $W\left(  E_{b}\right)  \in\mathbb{Z}$,
defined as \cite{PhysRevX.9.041015,PhysRevLett.124.086801}:  
\begin{equation}
W\left(  E_{b}\right)  =\int_{0}^{2\pi}\frac{dk}{2\pi i}\partial_{k}\ln\left(
h\left(  k\right)  -E_{b}\right),
\label{eq3}
\end{equation}
where $E_{b}\in\mathbb{C}$ is a reference point. If $W\left(  E_{b}\right)
\neq0$ for some $E_{b}$, the system under the OBC features
the non-Hermitian skin effect, which is a consequence of non-trivial
point-gap topology \cite{PhysRevLett.125.126402, PhysRevLett.124.086801}.

To obtain the OBC energy spectrum and quantify the non-Hermitian skin modes, we take the analytic continuation of $h\left(  k\right)$ by making the substitution $e^{ik}\rightarrow\beta=e^{\mu+ik}$ ($k$ and $\mu$ are real-valued). Then, for each $h\left(  k\right)  $, there exists a holomorphic
function
\begin{equation}
h\left(  \beta\right)  =t_{-M}\beta^{-M}+\cdots t_{N}\beta^{N}=\frac
{P_{M+N}\left(  \beta\right)  }{\beta^{M}},
\label{eq4}
\end{equation}
where $P_{M+N}\left(  \beta\right)  $ is a polynomial of order $M+N$. According to the fundamental theorem of algebra, for any $E\in\mathbb{C}$, the characteristic equation $f\left(  \beta,E\right) = h\left(  \beta\right)
-E=0$ has $M+N$ roots of $\beta$. We can order these roots in ascending amplitude $\left\vert \beta_{1}\left(  E\right) \right\vert \leq\left\vert \beta_{2}\left(  E\right)  \right\vert \leq
\cdots\leq\left\vert \beta_{M+N}\left(  E\right)  \right\vert $, and the GBZ is determined by the following equation \cite{PhysRevLett.121.086803,PhysRevLett.123.066404,PhysRevLett.125.126402}:
\begin{equation}
\left\vert \beta_{M}\left(  E\right)  \right\vert =\left\vert \beta
_{M+1}\left(  E\right)  \right\vert.
\end{equation}
All solutions of $\beta_{M},\beta_{M+1}$ trace a closed curve, termed the GBZ,
in complex $\beta$-plane. This curve encapsulates critical information about
eigenstate profiles, including the conventional wave vector $k$ and the
spatial decay rate $\mu$ of a skin mode. Upon obtaining the GBZ, we can acquire the OBC energy spectrum in the thermodynamic (large-size) limit by inserting $\beta\in $ GBZ
into Eq.~(\ref{eq4}). Furthermore, replacing the BZ with the GBZ allows us to define the
topological invariants that dictate the topological boundary modes, thereby
establishing the non-Bloch bulk-edge correspondence. Consequently, the GBZ serves a role analogous to the BZ in Hermitian systems and occupies a central position in the non-Bloch band theory.

A concept closely related to the GBZ is the aGBZ \cite{PhysRevLett.125.226402}, defined by projecting the following two equations onto the complex $\beta$-plane,%
\begin{equation}
f\left(  \beta,E\right)  =f\left(  \beta e^{i\theta},E\right)  =0,\theta
\in\mathbb{R}.
\label{eq6}
\end{equation}
Since there are five variables $\left(  \operatorname{Im}\beta
,\operatorname{Re}\beta,\operatorname{Re}E,\operatorname{Im}E,\theta\right)  $
and four constraint equations $\operatorname{Re}f=\operatorname{Im}%
f=\operatorname{Re}f_{\theta}=\operatorname{Im}f_{\theta}=0$, where
$f_{\theta}=f\left(  \beta e^{i\theta},E\right) $, the solution of Eq.~(\ref{eq6}) forms a 1D curve in the 5D space. The analytical
expression of aGBZ can be calculated via the resultant method to eliminate the
additional variables $\theta$ and $E$. The
resulting equation is an algebraic equation of $\operatorname{Re}\beta$ and
$\operatorname{Im}\beta$, given by%
\begin{equation}
\sum_{ij}c_{ij}\left(  \operatorname{Re}\beta\right)  ^{i}\left(
\operatorname{Im}\beta\right)  ^{j}=0.
\label{eq7}
\end{equation}
The aGBZ described by Eq.~(\ref{eq7}) is a complicated curve in complex $\beta$-plane,
composed of a set of analytic arcs connected at the self-intersection points. Specifically, each analytic arc consists of conjugate pair $\left(
\beta_{0},\tilde{\beta}_{0}\right)  $, where $\tilde{\beta}_{0}=\beta
_{0}e^{i\theta_{0}}$, satisfying $f\left(  \beta_{0},E_{0}\right)  =f\left(
\tilde{\beta}_{0},E_{0}\right)  =0$. Then, solving $f\left(  \beta
,E_{0}\right)  =0$ and ordering the roots by the absolute value, we can
identify the ordering of two roots that have the same absolute value as
$\left\vert \beta_{0}\right\vert $. For example, if\ $\left\vert \beta
_{0}\right\vert =\left\vert \beta_{M}\left(  E_{0}\right)  \right\vert
=\left\vert \beta_{M+1}\left(  E_{0}\right)  \right\vert $, the ordering of
$\beta_{0}$ is $(M,M+1)$, which helps us pick up the GBZ from the aGBZ. Any
points in an analytic arc have the same root ordering, changing only at the
self-intersection points.

\section{self-intersection in one-band non-Hermitian energy spectrum} \label{sec3}

For one-band energy spectra with non-trivial point-gap topology, the band structure can be categorized into two types: simple closed curves enclosing a single connected region or self-intersecting closed curves partitioning the interior into distinct sub-regions. To analyze the self-intersection points in the latter type of energy spectra, we define the following winding number
\begin{equation}
w\left(  \mathcal{C},E_{b}\right)  =\oint\nolimits_{\mathcal{C}}\frac{d\beta
}{2\pi i}\partial_{\beta}\ln[h\left(  \beta\right)  -E_{b}],
\label{eq8}
\end{equation}
where $\mathcal{C}$ is a closed contour in complex $\beta$-plane. Especially, if $\mathcal{C}$ is the unit circle (the BZ), then Eq.~(\ref{eq8}) reduces to Eq.~(\ref{eq3}).
According to the argument principle, the winding number of a complex
function is the difference between the total number of zeros and
poles enclosed by $\mathcal{C}$ \cite{ahlfors1979complex}, namely
\begin{equation}
w\left(  \mathcal{C},E_{b}\right)  =N_{1}-N_{2},
\end{equation}
where $N_{1} (N_{2})$ is the count of zeros (poles) with their respective multiplicities. As the contour $\mathcal{C}$ encloses the origin, we always have $N_{2}=M$. Therefore, the winding number $w\left(  \mathcal{C},E_{b}\right)  $ is determined by the number of zeros of $P_{M+N}\left(  \beta\right)  -\beta^{M}E_{b}$ that lie within the contour $\mathcal{C}$.

\begin{figure}[t]
    \includegraphics{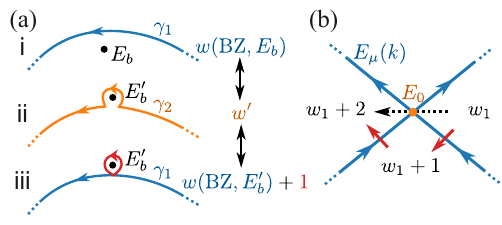}
  \caption{Relations of the winding numbers in distinct sub-regions. (a) Change of the winding number when the reference point passes across the PBC energy spectrum. $E_b$ and $E_b^{\prime}$ denote two different reference points, and the colored curves denote the loops in the complex energy plane. Panels i to iii show three equivalent loops with the same total winding number. (b) Winding numbers in the neighborhood of a self-intersection point. The blue lines denote the PBC spectrum, the orange point is the self-intersection point. The blue arrows denote the direction of $k$, and the winding number in the four sub-regions are marked in the figure.} 
  \label{fig:change-of-winding}
\end{figure}

Since the winding number defined in Eq.~(\ref{eq3}) is a topological number, it only changes when $E_b$ passes across the PBC spectrum, and the increment or decrement of the winding number is determined by the direction in which $E_b$ crosses the PBC spectrum. As illustrated in Fig.~\ref{fig:change-of-winding}(a), consider a reference point moving across a segment of the PBC spectrum [curve $\gamma_1$ in Fig.~\ref{fig:change-of-winding}(a)] 
 from the left-hand side ($E_b$) to the right-hand side ($E_b^{\prime}$), viewed from the direction in which $k$ increases. To calculate $w\left( \mathtt{BZ},E_{b}\right) - w\left( \mathtt{BZ},E_b^{\prime}\right)$, we continuously change the loop $\gamma_1$ into an auxiliary loop $\gamma_2$, shown as the panel ii in Fig.~\ref{fig:change-of-winding}(a). From i to ii, the reference point does not pass the loop, so that the winding number of the loop $\gamma_2$ around $E_b^{\prime}$, denoted as $w^{\prime}$, equals $w\left( \mathtt{BZ},E_{b}\right)$. Based on the auxiliary loop, we close the detour into a small loop (red loop in panel `iii'), and recover the PBC spectrum, then the winding number $w^{\prime}$ equals the sum of the two loops, i.e., $w\left( \mathtt{BZ},E_{b}\right) = w^{\prime}=w\left( \mathtt{BZ},E_b^{\prime}\right)+1$. As a result, when the reference point crosses the PBC spectrum from the left to the right, the spectral winding number decreases by $1$, and vice versa.

When the PBC spectrum is a self-intersecting closed curve, its interior is divided into distinct sub-regions, which is connected by the self-intersection points. 
As illustrated in Fig.~\ref{fig:change-of-winding}(b), we consider the simple crossing of the PBC spectrum at one intersection point $E_0$. The direction of $k$ is marked by the blue arrows. The crossing at the intersection point can be viewed as a node with four directed edges in graph theory. Because the PBC spectrum is closed loop on the complex plane, the number of edges towards the intersection point must equal the number of edges directed from the intersection point. Therefore, it is always possible for a reference point to cross the PBC  spectrum twice from the right to the left, as illustrated by the red arrows in Fig.~\ref{fig:change-of-winding}(b). Supposing the winding number around the start point is $w_1$, because each crossing will increase the winding number by $1$, so that the winding number around the end point equals $w_1+2$.  In the following, we demonstrate that such
a self-intersection point $E_0$ corresponds to the intersection points of the BZ curve and the aGBZ curve in the complex $\beta$-plane. We first
show that the self-intersection point  $E_0$ corresponds to the points on the unit
circle of the complex $\beta$-plane. Since 
 $E_{0}$  belongs to the PBC  spectrum, the characteristic equation
$f\left(  \beta,E_{0}\right)  =0$ must have two different solutions $\beta
_{0}^{\left(  1\right)  }\left(  E_{0}\right)  =e^{ik_{1}}$, $\beta
_{0}^{\left(  2\right)  }\left(  E_{0}\right)  =e^{ik_{2}}$, where
$k_{1},k_{2}\in(0,2\pi]$. Consequently, both $\beta_{0}^{\left(  1\right)  }$
and $\beta_{0}^{\left(  2\right)  }$ reside on the unit circle of complex
$\beta$-plane (namely the BZ curve).

Then, we show that $\beta_{0}^{\left(  1\right)  }$ and $\beta_{0}^{\left(
2\right)  }$ also lie on the aGBZ curve. In fact, the aGBZ curve can be
regarded as all possible solutions of $f\left(  \beta,E\right)  =0$
satisfying $\left\vert \beta_{i}\left(  E\right)  \right\vert =\left\vert
\beta_{j}\left(  E\right)  \right\vert $ for any $i\neq j$. Specifically, we
define a set of curves $\partial\mathcal{B}_{i,i+1}=\left\{  \beta
\in\mathbb{C}|\forall E\in\mathbb{C}:\left\vert \beta_{i}\left(  E\right)
\right\vert =\left\vert \beta_{i+1}\left(  E\right)  \right\vert \right\}  $
$\left(  1\leq i\leq M+N-1\right)  $, where the roots of $f\left(
\beta,E\right)  =0$ are ordered by their magnitude $\left\vert \beta
_{1}\left(  E\right)  \right\vert \leq\cdots\leq\left\vert \beta_{m+n}\left(
E\right)  \right\vert $. The aGBZ is the union of all these curves, that is,
$\mathtt{aGBZ}=\partial\mathcal{B}_{1,2}\cup\cdots\cup\partial\mathcal{B}_{M+N-1,M+N}$,
and the GBZ is just $\partial\mathcal{B}_{M,M+1}$. Each curve $\partial
\mathcal{B}_{i,i+1}$ forms a closed loop, enclosing $i$ zeros of the equation
$f\left(  \beta,E\right)  =0$ for arbitrary $E$ [see details in Appendix~\ref{appB}] . On the other hand, the
interior of the self-intersected PBC  spectrum is partitioned into several
sub-regions. Consider two sub-regions, labeled I and II, connected by the
self-intersection point $E_{0}$. For any reference point $E_{b1}$ in sub-region
I, the winding number along the BZ is
\begin{equation}
w\left( \mathtt{BZ},E_{b1}\right)  =l_{1}-M\equiv w_1,
\end{equation}
where $l_{1}$ is the number of zeros within the unit circle for the
characteristic equation $f\left(  \beta,E_{b1}\right)  =0$. Similarly, for
any reference point $E_{b2}$ in sub-region II, we have
\begin{equation}
w\left( \mathtt{BZ},E_{b2}\right)  =l_{2}-M \equiv w_2,
\end{equation}
where $l_{2}$ is the number of zeros within the unit circle for $f\left(
\beta,E_{b2}\right)  =0$. Note that $l_{1}\left(  l_{2}\right)  $ remains
invariant for any $E_{b1}\left(  E_{b2}\right)  $ within sub-region I (II).
As illustrated in Fig.~\ref{fig:change-of-winding}(b), there always exist two different regions around the self-intersection points that satisfy $l_{2}=l_{1}+2$ (i.e. $w_2=w_1+2$). The reference
points $E_{b1}$ and $E_{b2}$ can be connected through a path $\mathcal{S}$
that passes through $E_{0}$, with the remaining points of $\mathcal{S}$
belonging to sub-regions I and II. As $E_{b1}$ moves towards $E_{0}$ along the path $\mathcal{S}$, the distances between $E_{b1}$ and $E_{0}$ can become
arbitrarily small,  while there remain $l$ zeros within the BZ curve. Once
this reference point traverses $E_{0}$ and enters sub-region II, the number of zeros within the unit circle increases by two, resulting in $l_1+2$. Therefore, the
corresponding solutions for $f\left(  \beta,E_{0}\right)  =0$, $\beta
_{0}^{\left(  1\right)  }\left(  E_{0}\right)  $ and $\beta_{0}^{\left(
2\right)  }\left(  E_{0}\right)  $, are the boundary points separating the regions with $l_1$
and $l_1+2$ zeros within the BZ curve, and thus belong to $\partial
\mathcal{B}_{l_1+1,l_1+2}$. As a result, $\beta_{0}^{\left(
1\right)  }\left(  E_{0}\right)  $ and $\beta_{0}^{\left(  2\right)  }\left(
E_{0}\right)  $ are the intersection points of the BZ and the aGBZ sub-curve
$\partial\mathcal{B}_{l_1+1,l_1+2}$.

On the other hand, if the aGBZ intersects with the BZ, e.g., a point of the
aGBZ $\beta$ such that $\left\vert \beta\left(  E\right)  \right\vert =1$, the
definition of the aGBZ implies the existence of at least one distinct point
$\beta^{\prime}$ that also satisfies $\left\vert \beta^{\prime}\left(
E\right)  \right\vert =1$. Hence, we can express $\beta\left(  E\right)
=e^{ik}$ and $\beta^{\prime}\left(  E\right)  =e^{ik^{\prime}}$ with $k\neq
k^{\prime}$. This indicates that $E$ is a self-intersection point of the PBC 
energy spectrum.

\begin{figure}[t]
    \includegraphics{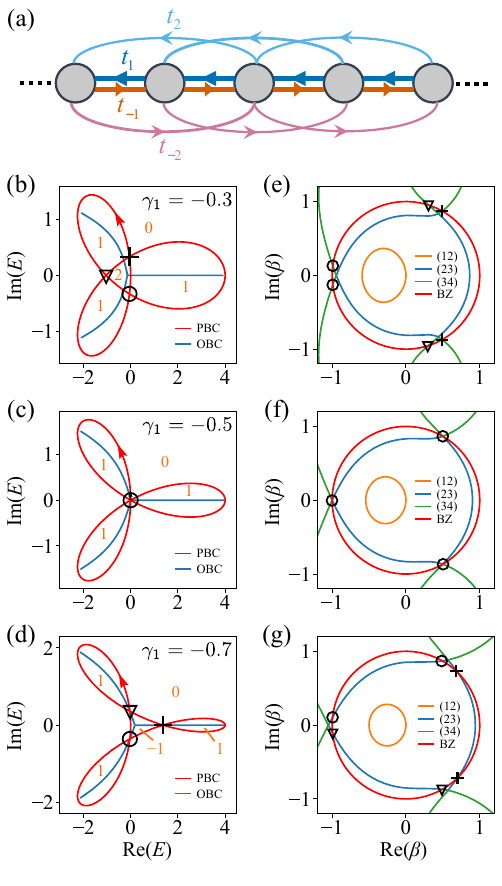}
  \caption{(a) Schematic of an extended Hatano-Nelson model with next-nearest neighbor hopping. (b)-(d) display the PBC (red line) and OBC (blue line) spectra of the model with varied parameters  $t_{1}=1+\gamma_1$ and $t_{-1}=1-\gamma_1$, where self-intersection points of the PBC energy spectra are marked by different shapes. Integer numbers in various regions denote winding numbers of the PBC energy spectra. (e)-(g) The corresponding aGBZs. Different colors represent different root orderings of the analytic arcs $\left\vert \beta_{i}\left(  E\right)  \right\vert =\left\vert
\beta_{i+1}\left(  E\right)  \right\vert $. Solutions for self-intersection points in complex $\beta$-plane are marked with the relevant shapes, denoting intersections between the aGBZ and the BZ. Throughout (b)-(g), we set $t_2=1.5$ and $t_{-2}=0.5$. } 
  \label{fig:1}
\end{figure}

To illustrate the above results, we consider an extended Hatano-Nelson
model with next-nearest neighbor hopping \cite{PhysRevLett.77.570,PhysRevB.56.8651}, shown in Fig.~\ref{fig:1}(a). The generalized Bloch Hamiltonian is given by%
\begin{equation}
h_{\mathtt{HN}}\left(  \beta\right)  =t_{-2}\beta^{-2}+t_{-1}\beta^{-1}%
+t_{1}\beta+t_{2}\beta^{2}\text{.}%
\label{eq12}
\end{equation}
If all hopping parameters are real, the real-space Hamiltonian $H_{\mathtt{HN}}$ obeys the generalized PT symmetry $\mathcal{K} H_{\mathtt{HN}} \mathcal{K} = H_{\mathtt{HN}}$, where $\mathcal{K}$ is the complex conjugate operator \cite{PhysRevLett.80.5243, CarlMBender_2002}.
The characteristic equation $h_{\mathtt{HN}}\left(  \beta\right)  -E=0$ has two poles and four zeros in complex $\beta$-plane. Therefore, the winding number
$w\left(  \mathcal{C},E_{b}\right)  $ for $h_{\mathtt{HN}}\left(
\beta\right)  $ along the curve $\mathcal{C}= \mathtt{BZ}$ can take integer values ranging from $-2$ to $2$. Furthermore, a self-intersection point of the PBC energy spectrum
correspond to two solutions of $\beta$ located on the unite circle. When
$t_{-2}=t_{2}=0$, this model reduces to the standard Hatano-Nelson model and the aGBZ is a circle with radius $\sqrt{\left\vert t_{1}/t_{-1}\right\vert }$
in the complex $\beta$-plane. For non-zero $t_{-2}$ and $t_{2}$, the aGBZ deviates
from a circle and can be obtained analytically using the resultant method
detailed in Appendix \ref{appA}. 

We present the energy spectra in Figs.~\ref{fig:1}(b)-(d) and their corresponding aGBZs
in Figs.~\ref{fig:1}(e)-(g), with $t_{2}=1.5$ and $t_{-2}=0.5$ fixed. The other two
parameters are set as $t_{1}=1+\gamma_1$ and $t_{-1}=1-\gamma_1$. When $\gamma_1=-0.3$, the interior of the PBC spectrum is divided into four distinct sub-regions:  three with a winding number of $1$ and one with a winding number of $2$ [see Fig.~\ref{fig:1}(b)]. In this scenario, the self-intersection points correspond to solutions of the characteristic equation satisfying $\left\vert \beta_{3}\left(  E\right)
\right\vert =\left\vert \beta_{4}\left(  E\right)  \right\vert =1$, which represent the intersection points between the GBZ and
the unit circle, as shown
in Fig.~\ref{fig:1}(e). For $\gamma_1=-0.7$, the interior of the PBC spectrum is similarly divided into four distinct sub-regions, three with a winding number of
$1$ and one with a winding number of $-1$ [see Fig.~\ref{fig:1}(d)]. The self-intersection points of the PBC spectrum, in this case, correspond to solutions of the characteristic equation satisfying $\left\vert
\beta_{2}\left(  E\right)  \right\vert =\left\vert \beta_{3}\left(  E\right) \right\vert =1$, as illustrated in Fig.~\ref{fig:1}(g). At $\gamma_1=-0.5$, all self-intersection points are merged at
zero energy, leading to three sub-regions, all with a winding number of $1$ [see Fig.~\ref{fig:1}(c)].
In this scenario, the self-intersection points satisfy the condition
$\left\vert \beta_{2}\left(  E\right)  \right\vert =\left\vert \beta
_{3}\left(  E\right)  \right\vert =\left\vert \beta_{4}\left(  E\right)
\right\vert =1$, as shown in Fig.~\ref{fig:1}(f). To analyze this condition, we consider
$\gamma_1=-0.5+\epsilon$, where $\epsilon$ is a small perturbation. For a
negative infinitesimal value of $\epsilon$, there exists a infinite small loop
with a winding number of $-1$,  admitting $\left\vert \beta_{2}\left(
E\right)  \right\vert =\left\vert \beta_{3}\left(  E\right)  \right\vert =1$
at the self-intersection points. Conversely, for a positive infinitesimal
$\epsilon$, a infinite small loop with a winding number of $2$ emerges,
resulting in $\left\vert \beta_{3}\left(  E\right)  \right\vert =\left\vert
\beta_{4}\left(  E\right)  \right\vert =1$ at these points. Therefore,
this triple self-intersection point corresponds to $\left\vert \beta
_{2}\left(  E\right)  \right\vert =\left\vert \beta_{3}\left(  E\right)
\right\vert =\left\vert \beta_{4}\left(  E\right)  \right\vert =1$ in complex
$\beta$-plane. Interestingly, the OBC spectra in Figs.~\ref{fig:1}(b)-(d) exhibit non-Bloch PT symmetry breaking \cite{PhysRevResearch.1.023013,Longhi:19,PhysRevLett.126.230402,lei2024activating}, where exceptional points arise. This type of non-Hermitian degeneracy corresponds to cusps of the GBZ curve in Figs.~\ref{fig:1}(e)-(g) \cite{PhysRevLett.132.050402} .

\section{self-intersection points in multi-band non-Hermitian energy spectrum}\label{sec4}

For a one-dimensional multi-band Hamiltonian, the characteristic equation is
now given by
\begin{equation}
f\left(  \beta,E\right)  =\det[h\left(  \beta\right)-E ]=0.
\label{eq12}
\end{equation}
This equation defines a two-dimensional Riemann surface in the
four-dimensional space $\left(  \operatorname{Re}\beta,\operatorname{Im}%
\beta,\operatorname{Re}E,\operatorname{Im}E\right) $. As $f\left(
\beta,E\right)  =\prod\nolimits_{\mu=1}^{n}\left[  E-E_{\mu}\left(
\beta\right)  \right]  =0$, each band $E=E_{\mu}\left(  \beta\right)  $
corresponds to a branch of the multi-valued function. Similar to the one-band case, the aGBZ is determined by the two roots of Eq.~(\ref{eq12}) with identical amplitudes. However, self-intersection points in multi-band cases can come from intersections between curves within the same sub-band or across different sub-bands. Therefore, the winding number should include the contributions from all sub-bands, namely
\begin{equation}
w\left(  \mathcal{C},E_{b}\right)  =\oint\nolimits_{\mathcal{C}}\frac{d\beta
}{2\pi i}\partial_{\beta}\ln\det[h\left(  \beta\right)  -E_{b}],
\end{equation}
where $\mathcal{C}$ is the BZ curve. Using the same method as in the
one-band case, it is obvious that self-intersection points in 
multi-band systems also arise from the intersection between the aGBZ and BZ curves in the complex $\beta$-plane.

\begin{figure}[t]
    \includegraphics[width=\columnwidth]{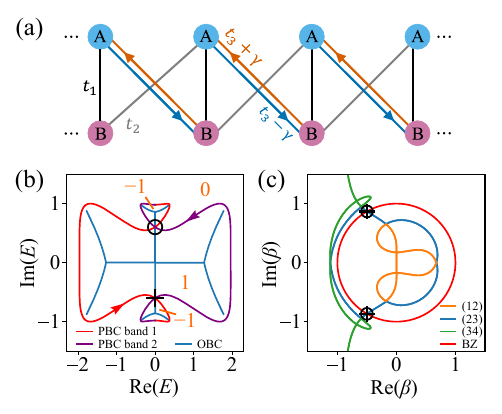}
  \caption{(a) The schematic of a non-Hermitian SSH model. (b) The PBC and OBC spectra in the complex energy plane. The self-intersection points of PBC energy spectra occur between two bands and are marked by different shapes. (c) The corresponding aGBZ in complex $\beta$-plane. Due to the chiral symmetry, the solutions for $\beta$ at self-intersection points are two-fold degenerate and denote the intersection of the aGBZ and the BZ. In (b) and (c), we set $t_1=t_2=\gamma=1$ and $t_3=0.2$.} 
  \label{fig:2}
\end{figure}

As an illustrative example, we consider a non-Hermitian SSH model \cite{PhysRevB.22.2099}, which is shown in Fig.~\ref{fig:2}(a). The generalized Bloch Hamiltonian of this two-band model reads
\begin{equation}
\begin{aligned}
    h\left(  \beta\right)  & = h_{x}\left(  \beta\right)  \sigma_{x}+h_{y}\left(
\beta\right)  \sigma_{y}; \\
h_{x}\left(  \beta\right) & = t_{1}+\frac{t_{2}+t_{3}+\gamma}{2}\beta
+\frac{t_{2}+t_{3}-\gamma}{2}\beta^{-1}, \\
h_{y}\left(  \beta\right) & = i\frac{t_{3}+\gamma-t_{2}}{2}\beta-\frac
{t_{3}-\gamma-t_{2}}{2}\beta^{-1},
\label{eq15}
\end{aligned}
\end{equation}
where $\sigma_{x,y,z}$ are the Pauli matrices. The absence of $\sigma_{z}$ indicates that this model possesses sublattice (or chiral) symmetry $\sigma_{z}^{-1}h\left(  \beta\right)  \sigma_{z}=-h\left(  \beta\right) $ \cite{Ryu_2010}, ensuring that the eigenvalues appear in pairs $\left(  E,-E\right) $, as shown in Fig.~\ref{fig:2}(b). The characteristic equation of Hamiltonian (\ref{eq15}) can be
written as
\begin{equation}
h_{x}^{2}\left(  \beta\right)  +h_{y}^{2}\left(  \beta\right)  -E^{2}=0,
\end{equation}
which is a quartic equation for $\beta$. Upon arranging the four solutions in ascending order of magnitude, denoted as $\left\vert 
\beta_{1}\right\vert
\leq\left\vert \beta_{2}\right\vert \leq\left\vert \beta_{3}\right\vert
\leq\left\vert \beta_{4}\right\vert $, the aGBZ corresponds to the trajectory of $\beta$ that satisfies the condition $\left\vert \beta_{i}\right\vert =\left\vert \beta_{j}\right\vert $. By numerically calculating the aGBZ in Fig.~\ref{fig:2}(c), we confirm that the self-intersection points of the PBC energy spectrum correspond to the intersections between the aGBZ and BZ curves.

\section{Theory of multiple self-intersection points}\label{sec5}

So far, we have primarily focused on the case of $2$-fold self-intersection
points, where the non-Hermitian energy spectrum crosses the self-intersection point only
twice. In this section, we extend our analysis to multiple self-intersection
points with multiplicities greater than two.

\begin{figure}[t]
    \includegraphics[width=\columnwidth]{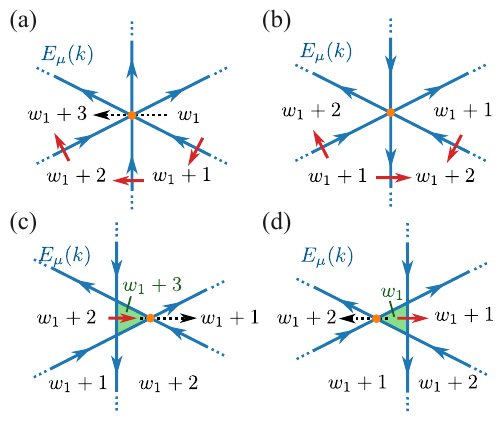}
  \caption{Panels (a) and (b) show the general two types of $3$-fold self-intersection points.  In the first type, the winding number can vary by up to 3 at the self-intersection point. In the second type, the change in winding number is limited to 1. However, small perturbations can split the second type into three double self-intersection points, generating a small loop with a winding number of $w_1$ in (c) or $w_1 +3$ in (d). Both types of $3$-fold self-intersection points lead to the condition $\left\vert \beta_{l+1}\left(
E\right)  \right\vert =\left\vert \beta_{l+2}\left(  E\right)  \right\vert
=\left\vert \beta_{l+3}\left(  E\right)  \right\vert =1$.   } 
  \label{fig:4}
\end{figure}

We first consider the $3$-fold self-intersection points in the PBC energy spectrum. These points, where the spectrum intersects itself thrice, connect six
distinct regions and can be categorized into two types. Without loss of
generality, the first type of $3$-fold self-intersection points connects regions
with a minimum winding number of $w_1=l-M$ and a maximum winding number of $w_1+3$
[see Fig.~\ref{fig:4}(a)]. Similar to the analysis of double self-intersection points,
we find that these points must satisfy $\left\vert \beta_{l+1}\left(
E\right)  \right\vert =\left\vert \beta_{l+2}\left(  E\right)  \right\vert
=\left\vert \beta_{l+3}\left(  E\right)  \right\vert =1$. The second type
connects regions with  a minimum winding number of $w_1+1$ ($l+1-M$) and a maximum winding number of $w_1+2$ [see Fig.~\ref{fig:4}(b)]. Introducing a small perturbation
transforms these $3$-fold self-intersection points into small loops. One case
occurs when the loop has a winding number of $w_1$ [see Fig.~\ref{fig:4}(c)],
resulting in the condition $\left\vert \beta_{l+1}\left(  E\right)
\right\vert =\left\vert \beta_{l+2}\left(  E\right)  \right\vert =1$ for these
self-intersection points. Another case arises when the loop has a winding
number of $w_1+3$ [see Fig.~\ref{fig:4}(d)], leading to the requirement that
$\left\vert \beta_{l+2}\left(  E\right)  \right\vert =\left\vert \beta
_{l+3}\left(  E\right)  \right\vert =1$. By combining these conditions, we
find that for both types of $3$-fold self-intersection points, $\left\vert
\beta_{l+1}\left(  E\right)  \right\vert =\left\vert \beta_{l+2}\left(
E\right)  \right\vert =\left\vert \beta_{l+3}\left(  E\right)  \right\vert =1$ always holds.

\begin{figure}[t]
    \includegraphics{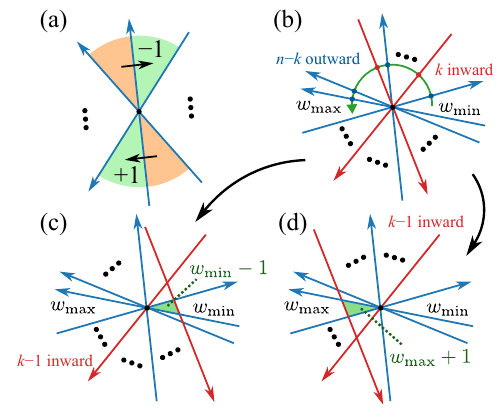}
  \caption{General proof for the $n$-fold self-intersection point. (a) Relation of the sum of winding number in opposite regions. (b) Illustration of the $n$-fold self-intersection point, where the green arc starts from the region with minimal winding number, rotating anticlockwise to the region with maximal winding number. The red and blue arrowed lines are trajectories of the PBC spectrum intersecting with the green arc at the half pointing to and away from the $n$-fold self-intersection point, respectively. (c) and (d) Two perturbations of one inward trajectory, where the winding numbers in the green triangular regions are marked by green font, respectively.} 
  \label{fig:general-proof}
\end{figure}

For general multiple self-intersection points with multiplicity $n$, we demonstrate that the $n$-fold self-intersection points of the spectrum correspond to $n$ solutions of $\beta$ on the BZ. The indices of these solutions are determined by the winding number in the vicinity of the $n$-fold self-intersection point, as well as the number of inward trajectories. In general, the neighborhood of an $n$-intersection point in the spectrum  is divided into $2n$ regions, each possessing distinct winding numbers. As illustrated in Fig.~\ref{fig:general-proof}(a), when two opposite regions (indicated by orange sectors) traverse along the same trajectory to adjacent opposite regions (depicted as green sectors), the increase in the winding number on one side is equal to the decrease on the other. Consequently, the sum of the winding numbers in opposite regions remains constant for a given
$n-$fold self-intersection point. Therefore, the regions with the maximum winding number $w_{\max}$ and the minimum winding number $w_{\min}$ are positioned opposite to each other. 

Consider the arc depicted in green in Fig.~\ref{fig:general-proof}(b), which originates from the region with $w_{\min}$ and rotates counterclockwise towards the region with $w_{\max}$. Let us assume that the arc intersects
$n-k$ trajectories extending outward from the
$n$-fold intersection point (indicated by blue arrowed lines) and
$k$ trajectories directed inward (shown with red arrowed lines). When the arc crosses an outward (inward) trajectory, the winding number increases (decreases) by $1$. Generally, the indices of the solutions for $\beta$ can be expressed as $\left|\beta_{l_{\text{min}}-k+1}\right|=\left|\beta_{l_{\min}-k+2}\right|=\cdots=\left|\beta_{l_{\max}+k}\right|=1$, where $l_{\min}=M+w_{\min}$ and $l_{\max}=M+w_{\max}$. This result can be demonstrated using a recursion method. For the $n$-fold self-intersection point,  if we pick one of the inward trajectories (represented by a red arrowed line), the multiplicity of the original self-intersection point is subsequently reduced by $1$.  For the perturbation shown in Fig.~\ref{fig:general-proof}(c), the winding number of the green region is $w_{\min}-1$, according to the relative position of the green region and the sector with $w_{\min}$. Conversely, for the perturbation shown in Fig.~\ref{fig:general-proof}(d), the winding number of the green region is $w_{\max} + 1$ for the analogous reason. We assume that our observation holds for $(n-1)$-fold self-intersection point. For the perturbation shown in Fig.~\ref{fig:general-proof}(c), since the maximal and minimal winding numbers of the $(n-1)$-fold self-intersection point are $w_{\max}$ and $w_{\min} - 1$, respectively, and the number of inward trajectories equals
 $k-1$, the relation $\left|\beta_{l_{\text{min}}-k+1}\right|=\left|\beta_{l_{\min}-k+2}\right|=\cdots=\left|\beta_{l_{\max}+k-1}\right|=1$ also holds. Similarly, for the perturbation shown in Fig.~\ref{fig:general-proof}(d), the relation $\left|\beta_{l_{\text{min}}-k+2}\right|=\left|\beta_{l_{\min}-k+3}\right|=\cdots=\left|\beta_{l_{\max}+k}\right|=1$ holds. By reducing the the amount of perturbation to $0$, both relationships will simultaneously hold for the $n$-fold self-intersection point, yielding  $\left|\beta_{l_{\text{min}}-k+1}\right|=\left|\beta_{l_{\min}-k+2}\right|=\cdots=\left|\beta_{l_{\max}+k}\right|=1$. Since the cases of $n=2$ and $n=3$ have been substantiated in our previous discussions, this conclusion is validated for arbitrary
$n$-fold self-intersection points through the recursion process. In Appendix \ref{appC}, we give a concrete method for constructing n-fold self-intersection points.

\section{summary and discussion}\label{sec6}

In this study, we unveil the geometric origins of self-intersection points in one-dimensional PBC energy spectra. We rigorously demonstrate that the self-intersection points of one-band PBC energy spectra arise from the intersection between the aGBZ and the BZ, which is verified numerically in an extended Hatano-Nelson model. Furthermore, we extend our analysis to multi-band cases, illustrated through a non-Hermitian SSH model. Moreover, we discuss the multiple self-intersection points, deriving the geometric conditions for general $n$-fold self-intersection points. We establish a one-to-multiple correspondence between non-Hermitian energy self-intersection points and the intersections between the aGBZ and the BZ. In general, for an $n$-fold self-intersection point $E_{0}$ of the PBC spectrum in the complex energy plane, the number of intersections, denoted as $\beta (E_{0})$, between the aGBZ and the BZ in the complex $\beta$-plane is equal to $n$.

When considering the case in which the dimension is greater than one, the
definitions of the aGBZ and the self-intersection points of the spectrum
appear ambiguous, as they involve multiple complex variables. Nonetheless, we can still apply our theory by analyzing one of the variables while holding the
others constant. For instance, in the two-dimensional case, the generalized
Bloch Hamiltonian for the one-band can be expressed as 
 $h\left(  \beta_{x},\beta_{y}\right)  =\sum_{m,n}c_{mn}\left(  \beta
_{x}\right)  ^{m}\left(  \beta_{y}\right)  ^{n}.$
The characteristic equation now takes the form $h\left(  \beta_{x},\beta
_{y}\right)  -E=0$, which is challenging to solve. However, for each fixed
$\beta_{x}$ or $\beta_{y}$, we can solve the characteristic equation similarly to the one-dimensional case, determining the aGBZ and the self-intersection points of the PBC energy spectrum.

As shown in Fig.~\ref{fig:1} and Fig.~\ref{fig:2}, when the winding numbers of the PBC spectra change sign, the corresponding self-intersection points are also located in the OBC spectra. This fact indicates that such self-intersection points serve as fixed points in the thermodynamic limit. In other words, as the system transitions from PBC to OBC by adjusting the boundary coupling \cite{PhysRevLett.127.116801,parto2025enhanced}, most PBC eigenstates transform into localized skin states, while the eigenstates associated with these self-intersection points remain extended. Additionally, the eigenstates corresponding to these self-intersection points exhibit scale-free properties in finite-size systems \cite{Li_2020,li2021impurity,PhysRevB.108.L161409, PhysRevB.107.134121, PhysRevB.109.035119}, thereby facilitating the experimental investigation of these special points.

\begin{acknowledgments}
  The authors would like to thanks Hui Tang, Guanghua Hua, and Qinxin Liu for their helpful discussion. We acknowledge financial support from the National Natural Science Foundation of China under Grant No. 12204395, Hong Kong RGC Early Career Scheme (Grant No. 24308323) and Collaborative Research Fund (Grant No. C4050-23GF), the Space Application System of China Manned Space Program, and Guangdong Provincial Quantum Science Strategic Initiative GDZX2404004. Yong-Chun Liu is supported by the National Key R \& D Program of China (Grant No. 2023YFA1407600), and the
  National Natural Science Foundation of China (NSFC)
  (Grants No. 12275145, No. 92050110, No. 91736106,
  No. 11674390, and No. 91836302).
\end{acknowledgments}

\appendix

\section{Calculation of aGBZ via the resultant method }\label{appA}

The calculation of aGBZ is the key to verify our results. In Appendix~\ref{appA}, we
give a brief introduction to the polynomial resultant theory and outline the
detailed application of this useful method to obtain the aGBZ.

In mathematics, the resultant of two polynomials is used to identify the
existence of a common factor, or equivalently, a common root. For two given
polynomials $f\left(  x\right)  =a_{n}x^{n}+\cdots+a_{0}$ and $g\left(
x\right)  =b_{m}x^{m}+\cdots+b_{0}$, the resultant with respect to the
variable $x$ is a polynomial expression of their coefficients, defined as
\begin{equation}
\mathtt{Res}_{x}\left[  f\left(  x\right)  ,g\left(  x\right)  \right]  =a_{n}^{m}%
b_{m}^{n}%
{\textstyle\prod\limits_{i,j}}
\left(  \xi_{i}-\eta_{j}\right) ,
\label{A1}
\end{equation}
where $\xi_{i}$ and $\eta_{j}$ are the root of $f\left(  x\right)  $ and
$g\left(  x\right)  $, respectively, with $1\leq i\leq n,1\leq j\leq m$.
Apparently, the polynomials $f\left(  x\right)  $ and $g\left(  x\right)  $
share a common root if and only if their resultant $\mathtt{Res}_{x}\left[  f\left(
x\right)  ,g\left(  x\right)  \right]  $ is zero. The resultant of $f\left(
x\right)  $ and $g\left(  x\right)  $ can be explicitly calculated using the
Sylvester matrix, defined as follows:%
\begin{equation}
\mathtt{Syl} \left(  f,g\right)  =\left[
\begin{array}
[c]{ccccccc}%
a_{n} & a_{n-1} & a_{n-2} & \cdots & 0 & 0 & 0\\
0 & a_{n} & a_{n-1} & \cdots & 0 & 0 & 0\\
\vdots & \vdots & \vdots &  & \vdots & \vdots & \vdots\\
0 & 0 & 0 & \cdots & a_{1} & a_{0} & 0\\
0 & 0 & 0 & \cdots & a_{2} & a_{1} & a_{0}\\
b_{m} & b_{m-1} & b_{m-2} & \cdots & 0 & 0 & 0\\
0 & b_{m} & b_{m-1} & \cdots & 0 & 0 & 0\\
\vdots & \vdots & \vdots &  & \vdots & \vdots & \vdots\\
0 & 0 & 0 & \cdots & b_{1} & b_{0} & 0\\
0 & 0 & 0 & \cdots & b_{2} & b_{1} & b_{0}%
\end{array}
\right]  .
\end{equation}
It can be proved that the resultant defined by Eq.~(\ref{A1}) is just the determinant of their Sylvester matrix \cite{lang2012algebra}, namely
\begin{equation}
\mathtt{Res}_{x}\left[  f\left(  x\right)  ,g\left(  x\right)  \right]  =\det
\mathtt{Syl}\left(  f,g\right).
\label{A3}
\end{equation}
Using this relation, we can derive the analytical expression of the aGBZ curve
on the complex $\beta$-plane by eliminating the variables $E$ and $\theta$ from
the equation
\begin{equation}
f\left(  \beta,E\right)  =f\left(  \beta e^{i\theta},E\right)  =0,\theta
\in\mathbb{R}.
\end{equation}
To eliminate $E$, we can calculate the resultant of $f\left(  \beta,E\right)  $
and $f\left(  \beta e^{i\theta},E\right)  $ with respect to $E$ from Eq.~(\ref{A3}), and
label this resultant as $G\left(  \beta,\theta\right)  ,$ namely%
\begin{equation}
G\left(  \beta,\theta\right)  =\mathtt{Res}_{E}\left[  f\left(  x\right)  ,g\left(
x\right)  \right]  =0.
\end{equation}
Since $G\left(  \beta,\theta\right)  $ is a complex algebraic function of
$\beta$ and $\beta e^{i\theta}$, the above equation can be separated into two
independent real algebraic equations as%
\begin{equation}
\operatorname{Re}G\left(  \beta,\theta\right)  =0,\operatorname{Im}G\left(
\beta,\theta\right)  =0
\end{equation}
Given that $\beta e^{i\theta}=\beta\cos\theta+i\beta\sin\theta,$ the variable
$\theta$ can not be directly eliminated using the resultant method. By
substituting $\cos\theta=\left(  1-t^{2}\right)  /\left(  1+t^{2}\right)  $
and $\sin\theta=2t/\left(  1+t^{2}\right)  $, we can obtain the resultant of
$\operatorname{Re}G\left(  \beta,t\right)  $ and $\operatorname{Im}G\left(
\beta,t\right)  $ with respect to $t$, leading to the expression for the
aGBZ:
\begin{equation}
F_{aGBZ}\left(  \operatorname{Re}\beta,\operatorname{Im}\beta\right)
=\mathtt{Res}_{t}\left[  \operatorname{Re}G\left(  \beta,\theta\right)
,\operatorname{Im}G\left(  \beta,t\right)  \right]  .
\end{equation}
The above result of $F_{aGBZ}\left(  \operatorname{Re}\beta
,\operatorname{Im}\beta\right)  $ is an algebraic polynomial of
$\operatorname{Re}\beta$ and $\operatorname{Im}\beta$, representing a closed
curve on the complex $\beta$-plane. When evaluating $F_{aGBZ}\left(
\operatorname{Re}\beta,\operatorname{Im}\beta\right)  $, symbolic computing
programs such as Mathematica and MATLAB can efficiently automate the
calculation of the resultant.

\section{the number of zeros enclosed by sub-aGBZ curves}\label{appB}
In Appendix~\ref{appB}, we show that each sub-curve $\partial\mathcal{B}_{i,i+1}$ is
the boundary of the open set that contains the $i$ zeros of the characteristic
equation. Consequently, $\partial\mathcal{B}_{i,i+1}$ is a closed curve that
encircles these $i$ zeros.

The single-band characteristic equation $h\left(  \beta\right)  -E=0$ establishes
a mapping from $\beta$ to $E$, indicating that for each $\beta$, there exists
a unique corresponding $E$.  Conversely, for a given value of $E$, there are
$m+n$ zeros in complex $\beta$-plane, where $m$ is the order of the pole.
These zeros can be ordered by their magnitude: $\left\vert \beta_{1}\left(
E\right)  \right\vert \leq\cdots\leq\left\vert \beta_{m+n}\left(  E\right)
\right\vert $. As $E$ sweep through the entire complex plane, we obtain a set
of continuum areas $\mathcal{A}_{1},\cdots,\mathcal{A}_{m+n}$. Here, the
definition of open set $\mathcal{A}_{i}$ is given by
\begin{equation}
\mathcal{A}_{i}=\left\{  \beta_{i} |\forall E\in\mathbb{C}%
:\left\vert \beta_{i-1}\left(  E\right)  \right\vert <\left\vert \beta
_{i}\left(  E\right)  \right\vert <\left\vert \beta_{i+1}\left(  E\right)
\right\vert \right\}  ,
\end{equation}
If there is an intersection between $\mathcal{A}_{i}$ and $\mathcal{A}_{j}$,
denoted as $I_{ij}$, then for any $\beta\in$ $I_{ij},$ it follows that
$\beta=\beta_{i}\left(  E\right)  =\beta_{j}\left(  E^{\prime}\right)  $ for
$E\neq E^{\prime}$. This implies that a single $\beta$ corresponds to two
distinct values, contradicting the injective nature of the function
$E=h\left(  \beta\right)  $. Therefore, $\mathcal{A}_{i}$ and $\mathcal{A}%
_{j}$ do not intersect in single-band systems; geometrically, this means that
$\mathcal{A}_{i}$ and $\mathcal{A}_{j}$ have no overlap.

The frontier of $\mathcal{A}_{i}$, denoted as $\partial\mathcal{A}_{i}$, is
the set of points $\beta$ such that every neighborhood of $\beta$ contains
both points in $\mathcal{A}_{i}$ and points not in $\mathcal{A}_{i}$.
Mathematically, the frontier of $\mathcal{A}_{i}$ can be expressed as
\begin{equation}
\partial\mathcal{A}_{i}=\left\{  \beta|\forall O\ni\beta:\mathcal{A}_{i}\cap
O\neq0,\mathcal{A}_{i}^{C}\cap O\neq0\right\}  .
\end{equation}
where $\mathcal{A}_{i}^{C}=\mathbb{C}\backslash\mathcal{A}_{i}$ is the
complementary set of $\mathcal{A}_{i}$. The boundary between $\mathcal{A}_{i}$
and $\mathcal{A}_{j}$ is defined as%
\begin{equation}
\partial\mathcal{B}_{i,j}=\partial A_{i}\cap\partial A_{j},
\end{equation}
Thus, we have%
\begin{equation}
\partial\mathcal{B}_{i,j}=\left\{  \beta_{i},\beta_{j}\in\mathbb{C}|\forall
E\in\mathbb{C}:\left\vert \beta_{i}\left(  E\right)  \right\vert =\left\vert
\beta_{j}\left(  E\right)  \right\vert \right\}
\end{equation}
and the GBZ is $\partial\mathcal{B}_{m,m+1}$. Furthermore, we define the set
$\mathcal{U}_{i,i+1}$ as the union of open sets
$\mathcal{A}_{1},\mathcal{A}_{2},\cdots,\mathcal{A}_{i}$ and their boundaries
$\partial\mathcal{B}_{r,s}$ $\left(  1\leq r,s\leq i\right)  $. The boundary of
$\mathcal{U}_{i,i+1}$ is the sub-curve\ $\partial\mathcal{B}_{i,i+1}$ of aGBZ.
According to the definition of $\mathcal{U}_{i,i+1}$, its interior
automatically contains the first $i$ zeros for any $E\in\mathbb{C}$. As a
result, the boundary $\partial\mathcal{B}_{i,i+1}$ forms a closed curve,
enclosing the $i$ zeros of the single-band characteristic equation $h\left(
\beta\right)  -E=0$.

\section{Constructing the $n$-fold self-intersection points.}\label{appC}

In the main text, we derive the geometric conditions for general n-fold
self-intersection points. Here, we present a straightforward and concrete
method for constructing n-fold self-intersection points. Taking the one-band
case as an example, we find that if the generalized Hamiltonian $h\left(
\beta\right)  $ can be expressed in the following form:%
\begin{equation}
h\left(  \beta\right)  =\left(  1-e^{i\phi}\beta^{n}\right)  q\left(
\beta\right)  ,
\end{equation}
where $q\left(  \beta\right)  $ is a holomorphic function of $\beta$ such that
$q\left(  \beta\right)  \neq0$ when $\left\vert \beta\right\vert =1$, and
$\phi$ is a real number satisfying $0\leq\phi<2\pi$, then $h\left(
\beta\right)  $ exhibits $n$-fold self-intersection points at the origin of
the complex energy plane. Specifically, the characteristic equation $h\left(
\beta\right)  -E=0$ for $E=0$ possesses $n$ distinct solutions that satisfy
$\left\vert \beta\right\vert =1$. These solutions are given by:
\begin{equation}
\beta=e^{i\frac{2\pi m-\phi}{n}},m=1,\cdots,n.
\end{equation}
For example, the generalized Hamiltonian in the Fig.~\ref{fig:1}(c) is
\begin{equation}
\begin{aligned}
h_{\mathtt{HN}}\left(  \beta\right)    & =0.5\beta^{-2}+1.5\beta^{-1}+0.5\beta
+1.5\beta^{2}\\
& =\left(  1+\beta^{3}\right)  \left(  0.5\beta^{-2}+1.5\beta^{-1}\right)  .
\end{aligned}
\end{equation}
In this case, we find $\phi=\pi$ and $q\left(  \beta\right)  =0.5\beta
^{-2}+1.5\beta^{-1}$. The solutions for this $3$-fold self-intersection point
are $\beta=e^{i\frac{\pi}{3}}$, $e^{i\pi}$, and $e^{i\frac{5\pi}{3}}$, which
is consistent with the numerical results shown in Fig.~\ref{fig:1}(f). 

\nocite{*}

\bibliography{apssamp}

\end{document}